\documentclass[%
 reprint,
 amsmath,amssymb,
 aps,
]{revtex4-1}

\usepackage{graphicx}% Include figure files
\usepackage{dcolumn}% Align table columns on decimal point
\usepackage{bm}% bold math
\begin{document}

\preprint{APS/123-QED}

\title{Effective Mass and Energy-Mass Relationship }% Force line breaks with \\
\thanks{Thanks to:  Prof. Shlomo Ruschin, Dr. Amir Natan of Tel Aviv University, and Dr. Leon Altschul of SDL Ltd. for reviewing this work.}%
\author{Viktor Ariel}
\affiliation{%
 Department of Physical Electronics\\
 Tel-Aviv University
}%

\begin{abstract}
The particle effective mass is often a challenging concept in  solid state physics due to the many different definitions of the effective mass that are routinely used. Also, the most commonly used  theoretical definition of the effective mass was derived from the assumption of a parabolic energy-momentum realtionship, $E(p)$, and therefore should not be applied to non-parabolic materials. In this paper, we use wave-particle duality to derive a definition of the effective mass and the energy-mass approximation suitable for non-parabolic materials. The new energy-mass relationship can be considered a generalization of Einstein's $E=mc^2$ suitable for arbitrary $E(p)$ and therefore applicable to solid state materials and devices.  We show that the resulting definition of the effective mass seems suitable for non-paraboic solid state materials such as  HgCdTe, GaAs, and  graphene. \\
\end{abstract}

\pacs{Valid PACS appear here}% PACS, the Physics and Astronomy
\maketitle

%\tableofcontents

\section{\label{sec:level1}Introduction\protect\\}
The concept of mass has a  long history in theoretical physics and philosophy \cite{Jammer} and continues to be hotly debated until today \cite{Okun}. In particular, in solid state materials many different effective mass definitions are used such as  the conductivity, density-of-states, and cyclotron effective mass \cite{Seeger}, \cite{Kittel}. The most commonly used theoretical definition of the effective mass is   $m=(\partial^2 E/ \partial p^2)^{-1}$, where $p=\hbar  k$ and $k$  is the crystal momentum \cite{Seeger}, \cite{Kittel}. We show that this definition is only  limited to parabolic $E(p)$ relationships while an alternative definition should be used for non-parabolic materials  \cite{ZawadzkiPf}, \cite{ArielThesis}, \cite{ArielPaper}. 

In this work, we use wave-particle duality by associating a particle with a one-dimensional wave-packet \cite{De Broglie}.  First, we derive a definition of the effective mass, $m$, and an approximation for the energy-mass relationship, $E(m)$. Then, we show that the new energy-mass relationship can be considered  a generalization of Einstein's $E=mc^2$ suitable for any dispersion $E(p)$ and therefore applicable to solid state materials and devices. Finally, we apply the definition of the effective mass to non-parabolic solid state materials  such as HgCdTe, GaAs, and graphene.

\section{Effective Mass Definition and  Energy-Mass Approximation}

We begin with  wave-particle duality by associating a particle with a wave-packet, and the particle velocity with the group velocity of the wave-packet, $v_g$.  This approach is commonly used in solid state physics for calculations of the energy-band structure and charge transport properties  \cite{Seeger}, \cite{Kittel}. Then based on the semi-classical definition of the particle momentum, the effective mass appears as a proportionality factor between the particle momentum and the group velocity of the wave-packet 
\begin{equation}
 p =    m v_g \ .
\label{eq:momentum}
\end{equation}
Consider a system of particles  described by an isotropic energy-momentum relationship $E(p)$.  Assume that the this $E(p)$ leads to a solution of the wave equation which can be represented by one-dimensional wave-packets. The group velocity is defined as the velocity of the wave-packet maximum and is usually approximated for one-dimensional packets as \cite{De Broglie}
\begin{equation}
 v_g \simeq  \frac {\partial E} { \partial  p} \ .
\label{eq:group}
\end{equation}
Similarly, we use the traditional one-dimensional approximations for the phase velocity, $v_p$, defined as the velocity of a single phase in the vicinity of the wave-packet maximum and expressed as \cite{De Broglie}
\begin{equation}
v_p\simeq \frac E  p \ .
\label{eq:phase}
\end{equation}

By using definitions of momentum  Eq.\ (\ref{eq:momentum}) and  group velocity  Eq.\ (\ref{eq:group}), we obtain the definition of the effective mass, which depends on the particle energy and momentum
\begin{equation}
 m  \equiv \frac {p} {\partial E / \partial p} \ .
\label{eq:mass}
\end{equation}
This result was previously demonstrated and applied to non-parabolic semiconductors  \cite{ZawadzkiPf}, \cite{ArielPaper}.

From Eq.\ (\ref{eq:momentum}) and Eq.\ (\ref{eq:phase}), we can derive a relationship between total partical energy, $E$, and particle effective  mass 
\begin{equation}
 E \simeq  m\   v_g  v_p \ .
\label{eq:mass-energy}
\end{equation}
Note from Eq.\ (\ref{eq:mass-energy}) that the energy-mass approximation is a function of both group and phase velocities of the corresponding wave packet. This seems like a remarkable relationship between wave and particle properties of matter within the limits of the present assumptions.

Also note that the above expressions   (\ref{eq:mass}) and  (\ref{eq:mass-energy}) are unique in a sense that no other one-dimensional definition of mass is possible without violating the present assumptions of the wave-particle duality  Eq.\ (\ref{eq:momentum}), and approximations for the group and phase velocities  Eq.\ (\ref{eq:group}) and  Eq.\ (\ref{eq:phase}). On the other hand,  it appears that  for any given dispersion relationship $E(p)$ , equations (\ref{eq:momentum}) - (\ref{eq:mass-energy}) comprise a complete  set of equations needed for description of the particle effective mass.  

For example, consider a special case of a free relativistic particle  described by the energy-momentum relationship $ E^2 = p^2 c^2 +m_0^2 c^4$, where $m_0$ is a constant rest-mass. By using this dispersion relationship in Eq.\ (\ref{eq:group}),  Eq.\ (\ref{eq:phase}), and  Eq.\ (\ref{eq:mass-energy}),  we obtain Einstein's energy-mass relationship
\begin{equation}
 E \simeq  mc^2 \ .
\label{eq:Einstein}
\end{equation}
It seems that the new approximation \ (\ref{eq:mass-energy}) can be considered  a generalization of Einstein's energy-mass relationship   \ (\ref{eq:Einstein})  for an arbitrary isotropic $E(p)$. 

Therefore,  we propose using the effective mass definition Eq.\ (\ref{eq:mass})  and the energy-mass approximation  Eq.\ (\ref{eq:mass-energy}) for particles with arbitrary $E(p)$  including non-parabolic  solid state materials.

\section{Application to non-parabolic solid state materials}

We would like to demostrate that the above definition of the effective mass Eq.\ (\ref{eq:mass}) and energy-mass approximation Eq.\ (\ref{eq:mass-energy})  can be successfully applied to carriers in non-parabolic solid state materials. 

First, consider the following theoretical definition of the effective mass which is often used  in solid state physics \cite{Kittel}, \cite{Seeger},
\begin{equation}
m=\frac {1} {\partial^2 E/ \partial p^2} \ .
\label{eq:mass-parab}
\end{equation}
This definition was derived based on the assumption of the constant energy-independent effective mass,  and therefore should not be used in non-parabolic $E(p)$, 
as we demonstrate in the Appendix. On the other hand, the effective mass defined by Eq.\ (\ref{eq:mass}) was not 
based on such an assumption and should be suitable for an arbtrary $E(p)$ relationship including non-parabolic solid state materials.

The electron kinetic energy in parabolic semiconductors, such as Si and Ge, near the conduction band minimum is approximated similar to the classical free particles
$  E \simeq { p^2} /{(2 m_0)}$,  where $m_0$ is assumed to be a constant coefficient.  
Then, for both the present definition of the effective mass Eq.\ (\ref{eq:mass})  and the parabolic definition  Eq.\ (\ref{eq:mass-parab}) we obtain 
\begin{equation}
m =  \frac {p} {\partial E / \partial p} =\frac {1} {\partial^2 E/ \partial p^2}= m_0 \ .
\label{eq:energy-par-semi}
\end{equation}

However, the situation is different in non-parabolic materials. 
Assume that the electron kinetic energy $E(p)$ in non-parabolic isotropic semiconductors, such as  HgCdTe and GaAs, 
can be described using a simplified one-dimensional Kane's approximation \cite{ZawadzkiPf} as 
\begin{equation}
 E +\alpha E^2 =\frac { p^2} {2 m_0} \ ,
\label{eq:energy-nonpar-semi}
\end{equation}
where the non-parabolicity factor, $\alpha$, and  constant mass, $m_0$, are considered energy independent coefficients.

Applying Eq.\ (\ref{eq:mass}),  we obtain the effective mass in non-parabolic semiconductors which is linearly dependent on energy \cite{ArielPaper}
\begin{equation}
m  = m_0\left (1+2\alpha E \right ) \ .
\label{eq:mass-nonpar-semi}
\end{equation}

Another example of a non-parabolic semiconductor is a  two-dimensional electron gas observed in graphene, 
which can be described by an isotropic relationship in the vicinity of the Dirac points \cite{Castro}  as 
 $E \sim v_f p $, where $v_f$ is the Fermi velocity. Using this dispersion relationship and definitions of the group and phase velocities  Eq.\ (\ref{eq:group}) and  Eq.\ (\ref{eq:phase}) leads to  $v_g = v_p=v_f $ in graphene.   Applying the energy-mass approximation  Eq.\ (\ref{eq:mass-energy}) we obtain 
\begin{equation}
 m \simeq  \frac {E} {v_g v_p}\ \simeq  \frac {p} {v_f}  .
\label{eq:ten}
\end{equation}
The above linear dependence of the particle effective mass on the momentum  was  confirmed by cyclotron resonance measurements in graphene \cite{Castro}. 
Note that the parabolic effective mass definition Eq.\ (\ref{eq:mass-parab})  leads to a mathematically divergent expression in graphene.   

Therefore, we demonstrated that the definition of the effective mass Eq.\ (\ref{eq:mass}) and the energy-masss approximation Eq.\ (\ref{eq:mass-energy}) 
are applicable to isotropic $E(p)$ in  non-parabolic solid statematerials such as HgCdTe, GaAs, and graphene.

\section{Conclusions}

In this work, we presented a simple  theoretical definition of the effective mass $m=p\cdot(\partial E/ \partial p)^{-1}$ and the energy-mass approximation $ E \simeq  m\   v_g  v_p$  based on the wave-particle duality and semi-classical definition of the particle momentum. The new energy-mass approximation is a function of  both group and phase velocities of the associated wave-packet  and can be considered a generalization of Einstein's $E=mc^2$  suitable for any $E(p)$ relationship. It appears that the  effective mass Eq.\ (\ref{eq:mass})  and the energy-mass approximation  Eq.\ (\ref{eq:mass-energy})  can be applied to non-parabolic solid statematerials such as HgCdTe, GaAs, and graphen. Therefore,  the effective mass definition and the energy-mass approximation seem suitable  for theoretical and experimental studies of particles with arbitrary dispersion relationships $E(p)$ where the wave-particle duality applies.

\pagebreak

\section{Appendix}

The parabolic definition of the effective mass Eq.\ (\ref{eq:mass-parab}) is typically derived assuming an approximate form of Newton's second law \cite{Kittel}, \cite{Seeger}, 
where the force, $F$, acting on a particle is approximated as
\begin{equation}
F=\frac {\partial p} {\partial t}\simeq m \frac {\partial v_g}{\partial t}\ .
\label{eq:newton-parab}
\end{equation}
Note that in (\ref{eq:newton-parab}) an assumption is made that mass is constant and not dependent on energy and time, which is the case for parabolic $E(p)$.

Then, using the definition of the group velocity   Eq.\ (\ref{eq:group}) we obtain

\begin{equation}
F \simeq m   \left(  \frac  {\partial}{\partial p}  \frac {\partial E}{\partial p} \right )  \frac  {\partial p}{\partial t}      \simeq m   \left(  \frac {\partial^2 E} {\partial \ p^2} \right )  \frac  {\partial p}{\partial t}   \ .
\label{eq:parab-derive}
\end{equation}

Comparing Eq.\ (\ref{eq:newton-parab}) and Eq.\ (\ref{eq:parab-derive}) leads to the traditional parabolic effective mass definition  Eq.\ (\ref{eq:mass-parab}).

Since the approximate form of Newton's second law Eq.\ (\ref{eq:newton-parab}) is only valid  for a parabolic $E(p)$, the resulting effective mass definition should not be used for non-parabolic materials.  When mass is  energy and time dependent, as in  non-parabolic $E(p)$,   the following exact form of Newton's second law should be used  \cite{Plastino} 
\begin{equation}
F=\frac {\partial p} {\partial t}= \frac {\partial m} {\partial t} v_g +m \frac {\partial v_g}{\partial t} \ .
\label{eq:newton-non-parab}
\end{equation}

It can be shown that using the exact form of Newton's second law Eq.\  (\ref{eq:newton-non-parab}) leads to the present effective mass definition  Eq.\ (\ref{eq:mass}).

\end{document}